\documentclass[fleqn,usenatbib]{mnras}

\usepackage{newtxtext,newtxmath}
\usepackage[T1]{fontenc}

\DeclareRobustCommand{\VAN}[3]{#2}
\let\VANthebibliography\thebibliography
\def\thebibliography{\DeclareRobustCommand{\VAN}[3]{##3}\VANthebibliography}

\usepackage{soul}
\usepackage{graphicx}	% Including figure files
\usepackage{amsmath}	% Advanced maths commands
\DeclareMathOperator{\sech}{sech}
\newcommand{\msun}{M$_\odot$}
\newcommand{\drvm}{$\Delta \textrm{RV}_\textrm{max}$}

\title[Galactic DWD population for \textit{LISA}]{Observationally driven Galactic double white dwarf population for \textit{LISA}}

\author[Valeriya Korol et al.]{
Valeriya Korol,$^{1}$\thanks{These authors contributed equally.}\thanks{E-mail: \href{mailto:korol@star.sr.bham.ac.uk}{korol@star.sr.bham.ac.uk} (VK)},
Na'ama Hallakoun$^{2}$\footnotemark[1] \thanks{E-mail: \href{mailto:naama.hallakoun@weizmann.ac.il}{naama.hallakoun@weizmann.ac.il} (NH)}, 
Silvia Toonen$^{1,3}$
and Nikolaos Karnesis$^{4,5}$
\\
$^{1}$Institute for Gravitational Wave Astronomy \& School of Physics and Astronomy, University of Birmingham, Birmingham, B15 2TT, UK \\
$^{2}$ Department of particle physics and astrophysics, Weizmann Institute of Science, Rehovot 7610001, Israel\\
$^{3}$ Anton Pannekoek Institute for Astronomy, University of Amsterdam, 1090 GE Amsterdam, The Netherlands\\
$^{4}$ APC, AstroParticule et Cosmologie, CNRS, Universite de Paris, F-75013 Paris, France\\
$^{5}$ Department of Physics, Aristotle University of Thessaloniki, Thessaloniki 54124, Greece
}

\date{Accepted 2022 January 31. Received 2021 December 17; in original form 2021 September 22}
\pubyear{2022}

\begin{document}
\label{firstpage}
\pagerange{\pageref{firstpage}--\pageref{lastpage}}
\maketitle

\begin{abstract}
Realistic models of the Galactic double white dwarf (DWD) population are crucial for testing and quantitatively defining the science objectives of the \textit{Laser Interferometer Space Antenna} (\textit{LISA}), a future European Space Agency's gravitational-wave observatory. In addition to numerous individually detectable DWDs, \textit{LISA} will also detect an unresolved confusion foreground produced by the underlying Galactic population, which will affect the detectability of all \textit{LISA} sources at frequencies below a few mHz.
So far, the modelling of the DWD population for \textit{LISA} has been based on binary population synthesis (BPS) techniques. The aim of this study is to construct an observationally driven population. To achieve this, we employ a model developed by Maoz, Hallakoun \& Badenes (2018) for the statistical analysis of the local DWD population using two complementary large, multi-epoch, spectroscopic samples: the Sloan Digital Sky Survey (SDSS), and the Supernova Ia Progenitor surveY (SPY). We calculate the number of \textit{LISA}-detectable DWDs and the Galactic confusion foreground, based on their assumptions and results.
We find that the observationally driven estimates yield 1) $2 - 5$ times more individually detectable DWDs than various BPS forecasts, and 2) a significantly different shape of the DWD confusion foreground. Both results have important implications for the \textit{LISA} mission. A comparison between several variations to our underlying assumptions shows that our observationally driven model is robust, and that the uncertainty on the total number of \textit{LISA}-detectable DWDs is in the order of 20\,per cent.
\end{abstract}


%%%%%%%%%%%%%%%%%%%%%%%%%%%%%%%%%%%%%%%%%%%%%%%%%%

%%%%%%%%%%%%%%%%% BODY OF PAPER %%%%%%%%%%%%%%%%%%

\section{Introduction}

Compact binaries and higher multiplicity systems composed of stellar remnants populating the local Universe are among the primary targets of the \textit{Laser Interferometer Space Antenna} (\textit{LISA}), an upcoming European Space Agency's gravitational-wave mission \citep[][]{lisa}, and the similar planned space-based gravitational-wave observatories \textit{TianQin} \citep{TianQin, hua20} and \textit{Taiji} \citep{Taiji}. Sensitive to gravitational waves in the $0.1 - 10$\,mHz frequency range, \textit{LISA} will detect various types of Galactic binaries with orbital periods extending up to a few hours \citep[e.g.][]{hil90,nel01,bre20}. Those composed of two white dwarfs (hereafter double white dwarfs, DWDs) will completely outweigh in number all the other Galactic and extra-galactic mHz-gravitational-wave sources. In particular, \textit{LISA} will detect the Galactic population of non-interacting (detached) DWDs in two ways: as individually resolved continuous quasi-monochromatic gravitational-wave signals, and as a persistent unresolved confusion foreground signal \citep[e.g.][]{cor02,far03,nel09}. The latter results from millions of overlapping monochromatic DWD signals that pile up at frequencies below a few mHz. The exact extent of the confusion foreground depends on the properties of the population, and on the \textit{LISA} mission duration \citep[e.g.][]{ben06,rui10,kar21}. Particularly, it is expected to be the dominant source of noise at frequencies lower than a few mHz, affecting the detectability of all the other \textit{LISA} sources, including merging massive black-hole binaries  \citep[$\sim 10^{4}$\,\msun\ $-10^{7}$\,\msun, e.g.][]{Klein2016}, extreme-mass-ratio inspirals \citep[e.g.][]{Babak2017,moo17,bon20}, and backgrounds of cosmological origin \citep[e.g.][]{Caprini2016,Tamanini2016}.
The modelling of the DWD population is therefore crucial for quantitatively defining \textit{LISA}'s science objectives, and for estimating the detection limits it imposes on other astrophysical sources \citep[e.g.][]{ada14,rob17,lit20,ant21,kar21,boi21}.

Characterising the Galactic DWD population with currently available electromagnetic facilities has proven to be technically challenging due to the compact size of these binaries and white dwarf stars, and their unique spectral characteristics \citep[e.g.][and see also Section~\ref{sec:obs} below]{reb19}. The discovery of DWDs that can be potentially detected by \textit{LISA} was made possible with specifically designed (electromagnetic) surveys (see Section~\ref{sec:obs}). As a result, the observed sample of \textit{LISA}-detectable DWDs is strongly biased and incomplete \citep[e.g.][]{kup18}, and all the current predictions for the \textit{LISA} mission are based on the binary population synthesis (BPS) technique rather than on actual observations \citep{nel01a,rui10,yu10,nis12,kor17,lam19,bre20,li20}. The BPS-based predictions are difficult to test against the observed sample \citep[but see][]{nel01a, too12} because of its limited size and nontrivial selection effects and biases (see Section~\ref{sec:obs}).

In this work, we step away from the BPS technique, and assemble a mock DWD population for \textit{LISA} based on results derived from observations. To achieve this, we follow the recipe for generating the Galactic disc DWD population of \citet[][]{maoz12} and the results obtained by fitting this model to observations in \citet{badenes12, maoz17} and \citet{maoz18}. We estimate a factor $2-5$ increase in the number of \textit{LISA} detections with this model compared to various BPS forecasts. Interestingly, the resulting population also produces a significantly different shape of the DWD confusion foreground compared to previous studies. Both results have important implications for Galactic as well as extra-galactic studies with the \textit{LISA} mission data.

The structure of this work is as follows. In Section~\ref{sec:obs} we give a brief overview of the current status of DWD observations without limiting ourselves to the sample of the currently known \textit{LISA} detectable binaries (also known as `\textit{LISA} verification binaries'). In Section~\ref{sec:pop} we describe how we construct the observationally motivated Galactic DWD population. In Section~\ref{sec:gwg} we provide a brief description of the \textit{LISA} data analysis method. In Section~\ref{sec:results} we present our results. In Section~\ref{sec:discussion} we discuss the implications for the \textit{LISA} observations compared to BPS models, and summarise the most important conclusions.

\section{Electromagnetic observations of DWDs} \label{sec:obs}

\subsection{The search for DWDs}

An efficient way to look for DWDs using electromagnetic observations is to search for radial velocity (RV) variations---shifts of up to a few hundred km\,s$^{-1}$ of a spectral line, induced by the orbital motion---in the optical spectra of WDs. However, the dense, compact, nature of WDs introduces a number of challenges for spectroscopic observations. Their high surface gravity induces gravitational settling, leaving only the lightest elements present in the atmosphere of the WD, so only hydrogen absorption lines are usually detected. The high density also induces significant pressure broadening on the absorption lines, making it much harder to obtain accurate RV measurements. The small size of WDs makes them intrinsically faint, so long exposure times and large telescopes are required in order to achieve a high enough signal-to-noise ratio (SNR) to enable high-resolution spectroscopy. On the other hand, since the orbital periods of close DWDs can be in the order of minutes to a few hours, long exposure times can smear the RV signal completely \citep[e.g.][]{reb19}.

The first systematic searches for detached close DWDs were carried out in the 1980s using the RV variation method, with limited success \citep{Robinson_1987, Foss_1991, Bragaglia_1990}. In 1992, \citet{Bergeron_1992} have identified WDs with masses so small ($\lesssim 0.45$\,\msun), they could not have had enough time to evolve from single stars, and must have undergone some phase of binary interaction in the past. This discovery has lead to a more fruitful approach of targeting low-mass WDs while searching for companions \citep[e.g.][]{Holberg_1995, Marsh_1995, Marsh_1995b, Maxted_2000}. Shortly after, \citet{Maxted_1999} have estimated a DWD fraction of $1.7-19$~per cent (with 95~per cent confidence), based on a sample of 46 WDs. By the beginning of the 2000s, only 18 close DWDs were known out of a total of $\sim 180$ WDs probed \citep{Marsh_2000, Maxted_2000}.

In the early 2000s, the Supernova Ia Progenitor surveY \citep[SPY;][]{Napiwotzki_2001} utilized the European Southern Observatory (ESO) Very Large Telescope (VLT) to obtain unprecedented high-resolution multi-epoch spectroscopy of a large sample of $\sim 1000$ relatively bright WDs. A few dozen DWD candidates were discovered in the SPY survey, some of which have been followed up and confirmed \citep[see][and references therein]{nap20}. The Sloan Digital Sky Survey \citep[SDSS;][]{York_2000} introduced a much larger, though less precise, multi-epoch spectroscopic sample of a few tens of thousands WDs, that has so far led to the discovery of a few DWDs and DWD candidates \citep[e.g.][]{Badenes_2009, Mullally_2009, badenes12, Breedt_2017, ved21}. Due to its lower resolution, the SDSS sample is sensitive only to DWDs with orbital separations $\lesssim 0.05$\,au. In the past decade, the Extremely Low Mass (ELM) survey of \citet{bro10}, designed to target $<0.3$\,\msun\ He-core WDs by applying a colour selection, has identified as many as 98 new DWDs \citep{Brown_2020}.

Although less efficient in terms of the required telescope time and the fraction of detectable systems, another method for finding DWDs is to look for eclipses, ellipsoidal modulations, or effects of companion irradiation, in photometric surveys. The first eclipsing DWD was discovered while searching for pulsating WDs \citep{Steinfadt_2010}. Among the $\sim 1000$ WDs observed by the \textit{Kepler K2} mission \citep{Howell_2014}, only one new DWD was discovered \citep[][along with one known DWD]{Hallakoun_2016, vanSluijs_2018}. The Zwicky Transient Facility \citep[ZTF;][]{Bellm_2019, Graham_2019, Masci_2019}, that has begun operating in 2018, shows promising results, with already $\sim 10$ newly discovered ultra-compact DWDs \citep[][and see also \citealt{vanRoestel21} for interacting DWDs]{Burdge_2019, Burdge_2020, Burdge_2020b, Coughlin_2020, Keller_2021}. In the near future, large all-sky surveys such as \textit{Gaia} \citep{Gaia_2016,Gaia_2018,Gaia_2020} and the Vera Rubin Observatory \citep{LSST} have the potential to deliver up to a few hundred and a few thousand of new eclipsing DWDs, respectively \citep{Eyer2012,kor17}. 

\begin{figure}
    \centering
    \includegraphics[width=\columnwidth]{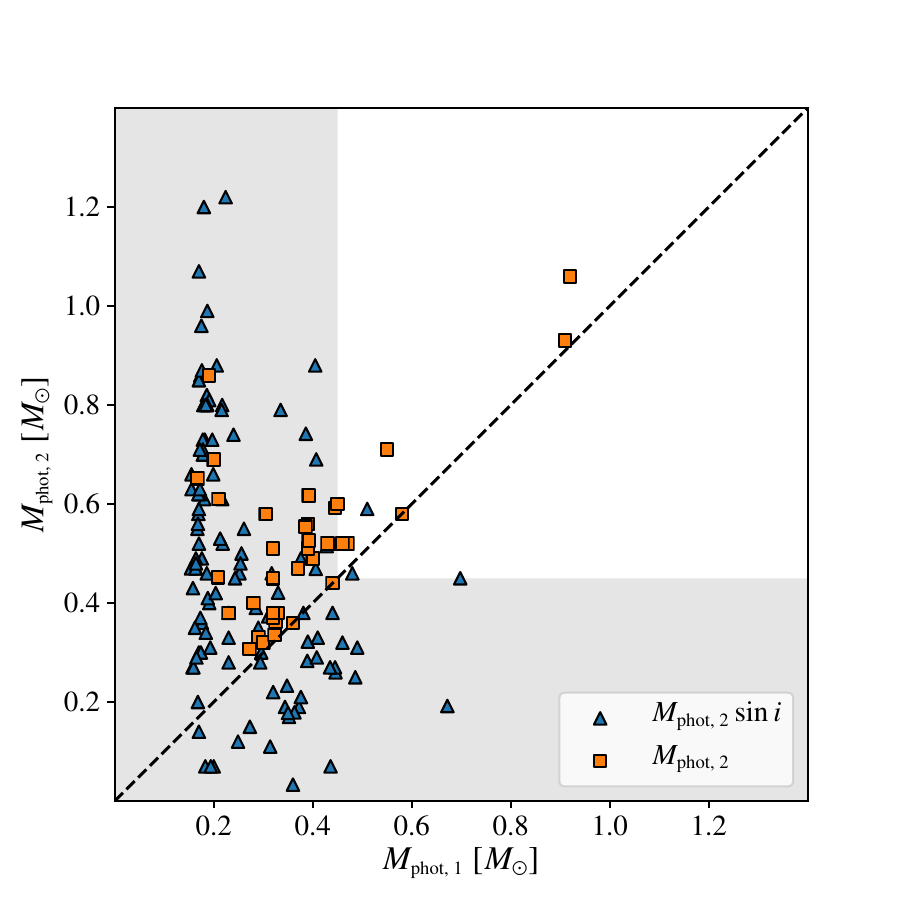}
    \includegraphics[width=\columnwidth]{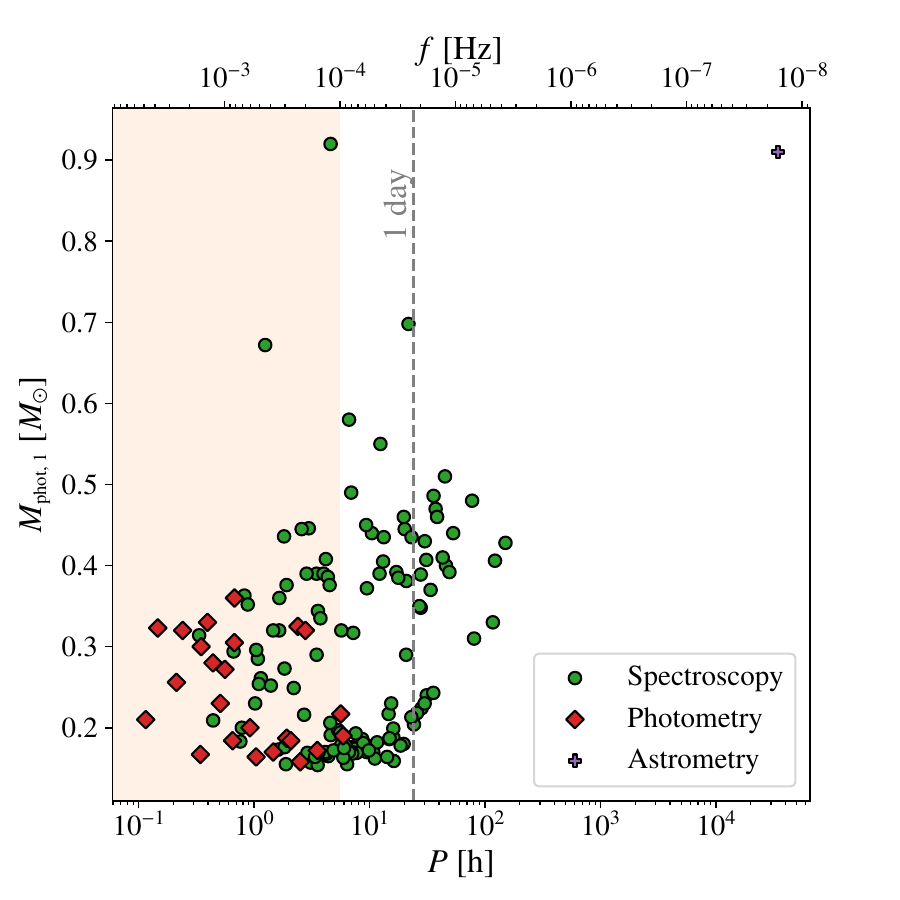}
    \caption{
    Known DWD systems parameters (the full reference list to appear in Hallakoun et al., in preparation).
    \textit{Top panel:} Component masses of the known DWD systems. Here $M_\textrm{phot, 1}$ is the mass of the photometric primary. Most of the systems (marked by blue triangles) have only a lower-limit estimate for the mass of the (unseen) companion, $M_\textrm{phot, 2}$. The known DWD sample is heavily-biased towards low-mass WDs ($<0.45$\,\msun, region highlighted in grey), which are easier to detect due to their higher luminosity and much higher binary fraction. \textit{Bottom panel:} Photometric primary mass as a function of the orbital period and \textit{LISA} frequency (top axis), of the known DWD systems. DWDs are marked by their detection method: spectroscopic RV variations only (green circles), photometric variations (red diamonds), and astrometry (purple plus shape). The orange-shaded area highlights the frequency range detectable by \textit{LISA}. There is a clear observational bias against detecting systems with orbital periods longer than 1\,day (indicated by the grey vertical dashed line).} 
    \label{fig:KnownDWDs}
\end{figure}

To date, the DWD sample amounts to $\sim 150$ systems with known orbital parameters, most of them include a low-mass WD (about 60~per cent of the systems include an ELM WD with mass $<0.25$\,\msun, and almost 95~per cent of the systems include a WD with mass $<0.45$\,\msun). Figure~\ref{fig:KnownDWDs} shows the masses and orbital periods measured for the known DWDs. Albeit being highly efficient in detecting DWDs, the ELM survey probes only the low-mass range of the DWD population. Our current knowledge of the physical characteristics of the general DWD population is thus rudimentary, at best. 

\subsection{DWD population properties based on the SDSS and SPY samples}

\citet{maoz12} have constructed a method for statistically characterising the binarity of a stellar population using sparsely sampled RV data of a large sample of stars. This is achieved by measuring \drvm, i.e. the maximal RV shift observed between repeated observations of the same star. The shape of the \drvm\ distribution of the full sample can then constrain the binary fraction of the population within the maximal separation to which the data are sensitive, along with the shape of the binary separation distribution. \citet{badenes12} applied this method to a sample of $\sim 4000$ WDs from the SDSS, and estimated that $3-20$~per cent of the WDs are in fact DWDs. Assuming a power-law separation distribution at birth (after the last common-envelope phase, see Section~\ref{sec:DWD_separations}), they estimated a slope between $-2$ and $+1$. These rather large uncertainties resulted from the large measurement errors of the SDSS sample, which also limited the sensitivity of the sample to orbital separations smaller than $\sim 0.05$\,au.

Utilising its higher precision, \citet{maoz17} applied the \drvm\ method to a subsample of $439$ WDs from the SPY survey. They estimated a binary fraction of $\sim 10$~per cent, with a power-law index $\sim -1.3$ for the orbital separation distribution. The smaller measurement errors also improved the sensitivity to orbital separations as large as $\sim 4$\,au. On the other hand, because of its smaller sample size, SPY was less likely to find rare ultra-compact DWDs, making the two samples complementary.
Later, \citet{maoz18} combined the results from both surveys---the SDSS and SPY---to obtain improved constraints on the properties of the DWD population. They estimated that a fraction of $0.095 \pm 0.020$ ($1\sigma$, random) $+0.010$ (systematic) of the WDs are DWDs with orbital separations $\lesssim 4$\,au, and that the index of the power-law distribution of the initial DWD separations is $\alpha = -1.30 \pm 0.15$ ($1\sigma$) $+0.05$ (systematic).
In the following we use these results to estimate the Galactic DWD population observable by \textit{LISA}.

\section{Assembling a mock population for \textit{LISA}} \label{sec:pop}

The gravitational-wave radiation from a DWD can be considered as a quasi-monochromatic signal that can be fully described by eight parameters: frequency $f$, frequency derivative or chirp $\dot{f}$, amplitude ${\cal A}$, sky position in ecliptic coordinates $(\lambda,\beta)$, orbital inclination $\iota$, polarisation angle $\psi$, and initial orbital phase $\phi_0$ \citep[e.g.][]{cut98,roe20,kar21}. In this section, we describe the recipes for generating ${\cal A}$, $f$, $\dot{f}$, and $(\lambda,\beta)$ based on the currently available observations. The binary inclination is drawn from a flat distribution in $\cos{\iota}$, while the polarization and the initial phase are chosen randomly from a flat distribution.

The gravitational-wave frequency of a quasi-monochromatic signal is given by
\begin{equation}
    f = \frac{2}{P},
\end{equation}
where $P$ is the DWD's orbital period.
The chirp of a binary is defined as
\begin{equation}\label{eq:fdot}
\dot{f}=\frac{96}{5}  \pi^{8/3} \left( \frac{G{\cal M}}{c^3} \right)^{5/3} f^{11/3},
\end{equation}
where $G$ and $c$ are the gravitational constant and the speed of light, respectively, and
\begin{equation} \label{eq:chirp_mass}
{\cal M}=\frac{(m_1 m_2)^{3/5}}{(m_1+m_2)^{1/5}},
\end{equation}
is the so-called chirp mass, with $m_1$ and $m_2$ being the primary and secondary WD masses. 
Given $f$, $\cal{M}$, and the luminosity distance to the binary $d$, one can determine the amplitude of the gravitational-wave signal:
\begin{equation}\label{eq:amp}
    {\cal A} = \frac{2(G{\cal M})^{5/3}(\pi f)^{2/3}}{c^4 d}.
\end{equation}
In the following subsections we describe the observationally motivated distributions that we use to sample $m_1$, $m_2$, $P$, and the disc density distribution that defines DWDs' 3D coordinates $(\lambda, \beta, d)$.

\subsection{White dwarf masses} \label{sec:wd_masses}

We assume that the primary mass, $m_1$---defined here as the heaviest of the two---follows the mass distribution of single WDs. Based on the SDSS spectroscopic sample of $\sim1500$ WDs, \citet{kep15} derived a WD mass function that follows a three-component Gaussian mixture with means $\mu = \{0.65, 0.57, 0.81\}$\,\msun\,, standard deviations $\sigma = \{0.044, 0.097, 0.187\}$\,\msun\,, and respective weights $w = \{0.81, 014, 0.05\}$. We note, however, that for close DWDs it is natural to expect that the more massive primary WD, which had evolved off the main sequence first, could have been affected by mass transfer and mass-loss processes. Therefore, it is likely that the primary mass distribution in DWDs is different compared to that of single WDs. We explore different variations to this assumption in Section~\ref{sec:model_variations} below.

Observations show that at the main-sequence stage, the secondary stars follow a mass-ratio distribution that is approximately flat, rather than the same mass distribution as primary stars \citep{rag10,duc13, Moe_2017}.
It is also unlikely that later, at the DWD evolution stage, the mass of the secondary WD will follow the same distribution as that of the primary.
At any orbital period, the number of observed DWD systems for which both masses can be determined is still too low, and the sample is too biased, to infer the mass distribution in a statistically robust way. 
The \drvm\ distribution analysed in \citet{maoz12} depends only weakly on the $m_1/m_2$ ratio and, thus, cannot constrain the mass-ratio distribution.
Therefore, in our default model we draw $m_2$ from a flat distribution between $0.15$\,\msun, the minimum mass of observed ELM WDs, and $m_1$.
For the small number of cases in which we get $m_1 < 0.25$\,\msun, i.e. falling into the ELM category, we draw $m_2$ from the range $[0.2, 1.2]$\,\msun\ with equal probability. Note that in the later case the definition of $m_1$ and $m_2$ is swapped. However, since the gravitational wave radiation depends only on the combined chirp mass (Eq.~\ref{eq:chirp_mass}), this does not affect our analysis.

\subsection{DWD separation distribution} \label{sec:DWD_separations}

Central to this work is the use of the DWD separation distribution inferred by \citet{maoz18} by combining the SDSS and SPY samples. Here we briefly summarise their assumptions and results.

The present-day DWD separation distribution can be derived analytically \citep[for details see][]{maoz12}, starting with the assumption that at the time of the DWD formation (i.e. after the final common-envelope phase), the separation distribution follows a power-law
\begin{equation}
n \left( a \right) \propto a^\alpha,
\end{equation}
which is similar to the initial zero-age main-sequence separation distribution \citep[with $\alpha \sim -1$;][]{Abt_1983}. Although the actual separation distribution might be more complicated, given that close DWDs have gone through multiple mass-transfer and/or common-envelope phases \citep[e.g.][]{nel01,rui10, too12}, in the limited orbital separation range considered here a power-law serves as a good approximation for other types of monotonic relations. This assumption is supported by the Type-Ia supernova delay-time distribution (DTD)---the time between the formation of the binary and the merger leading to the Type-Ia explosion. Observations have shown that the DTD approximately follows a $\propto t^{-1}$ relation \citep[see e.g.][for a review]{Maoz_2014}. Theoretical studies have demonstrated a strong dependence of the DTD on the initial DWD separation distribution, where a $\sim t^{-1}$ DTD is the generally expected form for the gravitational-wave-induced mergers of a single-age DWD population with a power-law initial separation distribution \citep[e.g.][]{Greggio_2005, Totani_2008}. This is also the DTD generally predicted by BPS models for the double-degenerate channel \citep[e.g.][]{too12}. The similarity between the observed Type-Ia supernova DTD and the expected gravitational-wave-induced merger rate of DWDs is considered as a strong argument in support of the DWD merger formation scenario of Type-Ia supernovae \citep[e.g.][]{Maoz_2014}.

Next, assuming a constant star formation history and setting the age of the Milky Way disc stellar population to $t_0 =10$\,Gyr \citep[e.g.][]{mig21}, the present-day DWD separation distribution is given by 
\begin{equation} \label{eqn:sep}
N  \left(x\right) \propto 
\begin{cases}
x^{4+\alpha}\left[ \left(1 + x^{-4} \right)^{\frac{\alpha+1}{4}} - 1 \right] &  \text{for} \ \alpha \neq -1, \\
x^3 \ln{\left(1+x^{-4} \right)} & \text{for} \ \alpha=-1
\end{cases}
\end{equation}
where $x \equiv {a}/{\left( Kt_0 \right)^{1/4}}$
is the separation normalised to the separation at which the binary will merge within $t_0$, and 
\begin{equation}
K\equiv \frac{256}{5}\frac{G^3}{c^5} m_1 m_2 \left( m_1 + m_2\right).
\end{equation}
Note that Eq.~\eqref{eqn:sep} is approximately a broken power-law with an index $\alpha$ when $x \gg 1$ (i.e. when the DWD merger time is greater than the age of the Galactic disc). When $x \ll 1$, the power-law index is 3 for $\alpha \ge -1$, and $\alpha+4$ for $\alpha \le -1$. 
For each simulated DWD system, $a$ is drawn from the $N(x)$ distribution, between $a_{\rm min} = 2 \times 10^4\,$km (DWD at contact) and $a_{\rm max}$.
We set $a_{\rm max}$ to the orbital separation corresponding the \textit{LISA}'s frequency lower limit of $1 \times 10^{-4}\,$Hz.

\subsection{DWD spatial distribution} \label{sec:DWD_spatial_distribution}

We assume that WDs (including DWDs) are distributed in the disc according to an exponential radial stellar profile with an isothermal vertical distribution
\begin{equation} \label{eqn:dwd_positions}
\rho(R,z)= \rho_{{\rm WD},\odot} \, e^{-\frac{R-R_\odot}{R_{\rm d}}} \sech^2 \left( \frac{z-z_\odot}{z_{\rm d}} \right),
\end{equation}
where $\rho_{{\rm WD},\odot} = (4.49 \pm 0.38) \times 10^{-3}$ pc$^{-3}$ is the local WD density estimated by \citet{hol2018} based on the 20\,pc WD sample in the second \textit{Gaia} Data Release \citep{Gaia_2018}, $0 \le R \le 20$\,kpc is the cylindrical radial coordinate measured from the Galactic centre, $z$ is the height above the Galactic plane, $R_{\rm d} = 2.5\,$kpc is the disc scale radius, and  $z_{\rm d}=0.3$\,kpc is the disc scale height \citep[e.g.][]{Juric2008,mac17}. The chosen values for scale parameters have been derived for red giant stars, i.e. WD progenitors. Because WDs are born with no recoil kick, it is reasonable to expect that DWDs should follow the density distribution of their progenitor population. Although a small recoil has been hinted at in wide DWD binaries \citep{El-Badry2018}, it does not affect the shortest period binaries considered in this work. Finally, we set the Sun's position to $(R_\odot,z_\odot) = (8.1,0.03)$\,kpc \citep[e.g.][]{abu19}.

\subsection{DWD total number} \label{sec:DWD_tot}

Finally, we estimate how many DWDs to draw from the above distributions. 
First, we calculate the total number of WD stars in the Milky Way disc by integrating the WD density profile given in Eq.~\eqref{eqn:dwd_positions}, which yields $(2.7 \pm 0.2) \times 10^{9}$ WDs, propagating the uncertainty of $\rho_{\rm WD, \odot}$ estimated by \citet{hol2018}.
To arrive to the total number of DWDs in the \textit{LISA} band, we have to multiply this number by the DWD fraction, $f_{\rm DWD} = 0.095 \pm 0.020$, derived by \citet{maoz18} for orbital separations $\lesssim 4$\,au. 
We re-scale this fraction to the maximum DWD separation detectable by \textit{LISA}, $a_\textrm{max}$, using
\begin{equation}
\label{eq:scaled_fbin}
f_{\textrm{DWD, }a_\textrm{max}} = \frac{\int_{a_\textrm{min}}^{a_\textrm{max}}N\left(a, \alpha \right) da}{\int_{a_\textrm{min}}^{4\,\textrm{au}}N\left(a, \alpha \right) da} f_{\textrm{DWD, 4au}},   
\end{equation}
where $a_{\rm min} = 1.3\times 10^{-3}$\,au (i.e. $2\times 10^4\,$km) set to be the same for all binaries, while $a_{\rm max}$ is the separation corresponding to the \textit{LISA}'s frequency lower limit of $1 \times 10^{-4}\,$Hz computed for each binary based on the binary's components masses using Kepler's law (see Section~\ref{sec:DWD_separations}).
Given that $\int N \left( a, \alpha \right) da = \frac{1}{\left( Kt_0 \right)^{1/4}}\int N \left( x, \alpha \right) dx$, we can integrate over $x$:
\begin{multline}
\int_{x_1}^{x_2} N \left(x\right) dx \propto\\
\begin{cases}
\left.\frac{x^\alpha}{\alpha + 5} \left[ x \left(1 + x^4 \right)\left(1 + x^{-4}\right)^{\frac{\alpha+1}{4}} - x^5 \right]\right|_{x_1}^{x_2}   & \text{for} \ \alpha \neq -1, \\
\left. \frac{1}{4} \left[ x^4 \ln \left( 1 + x^{-4} \right) + \ln \left( 1 + x^4 \right) \right] \right|_{x_1}^{x_2} & \text{for} \ \alpha=-1.
\end{cases}
\end{multline}
Note that the maximum separation detectable by \textit{LISA}, $a_{\rm max
}$, depends on the total mass of the binary through Kepler's law.
For example, \textit{LISA}'s low-frequency limit of 0.1\,mHz corresponds to an orbital separation of $\sim$0.0049\,au for a 0.6\,\msun+0.6\,\msun\ DWD, which decreases to $\sim0.0043$\,au for a 0.6\,\msun+0.2\,\msun\ DWD. 
Therefore, we sample a statistically significant number of DWD masses, as described in Section~\ref{sec:wd_masses}, and compute $f_{\rm DWD}$ each time.
We then take the median of the resulting distribution as representative of the entire population.
The obtained median $f_{\rm DWD}=0.010 \pm 0.003$ yields $(26 \pm 6) \times 10^6$ DWDs in the \textit{LISA} frequency band, where the reported error accounts for the uncertainties of $\rho_{\rm WD, \odot}$ and $f_{\rm DWD}$. Only a small fraction of this total number is expected to be individually resolved, while the remaining majority will contribute to the unresolved confusion signal (see Section~\ref{sec:results}). Note that our estimate of the DWD total number has been obtained using the best-fit value of $\alpha=-1.3$ derived in \citet{maoz18}. This value implies a relatively flat separation distribution in logarithmic scale, consistent with BPS simulations, and with a DTD of $\sim t^{-1}$ (see Section~\ref{sec:DWD_separations}). Varying $\alpha$ within the 1$\sigma$ confidence interval results in $\pm 10\,$per cent variation in the DWD total number. 

We highlight that our estimate of the DWD total number scales with the Galactic disc volume, the WD local density $\rho_{\rm WD,\odot}$, and the DWD fraction $f_{\rm DWD}$. Importantly, this also means that our results (see Section~\ref{sec:results}) can be re-scaled to desired values of $\rho_{\rm WD,\odot}$ and $f_{\rm DWD}$ when these values will be better constrained by new observations and larger samples.

\subsection{Model variations} \label{sec:model_variations}

In addition to the default model assumptions described in Sections~\ref{sec:wd_masses} and \ref{sec:DWD_separations}, we compile eight alternative DWD model variations by changing the primary and secondary mass distributions, the ELM mass limit, and the adopted values for $\alpha$ and $f_{\rm DWD}$. We motivate these below.

First, we investigate the effect of changing the primary mass distribution. Our default assumption is based on the mass distribution of single WDs, in the absence of a statistically reliable measurement of the primary mass distribution of WDs in binaries. However, the actual primary mass distribution of close DWDs is likely to be different than that of single WDs, given the mass transfer between the binary components during their past common-envelope phases. We thus experiment by manually shifting the mean mass of the primary Gaussian of the \citet{kep15} distribution (see Section~\ref{sec:wd_masses}) by $\pm 0.1$\,\msun\ from the default value of 0.65\,\msun, while maintaining the shape of the other two Gaussian components (`Primary Gaussian+0.1\,M$\odot$' and `Primary Gaussian$-$0.1\,M$\odot$' model variations hereafter). Alternatively, we replace the primary mass distribution of \citet{kep15} by the observed distribution of the systems that constitute the `core' of the \drvm\ distribution of the SPY sample (consisting of mostly single WDs), using a three-component Gaussian mixture with $w = \{0.66,0.24,0.1\}, \mu = \{0.55,0.72,0.3\}$\,\msun, and $\sigma = \{0.07,0.3,0.078\}$\,\msun\ \citep[][`Primary SPY core' model-variation]{maoz17}.

Second, we explore the impact of changing the secondary mass distribution. Since the visible component in single-lined DWDs is usually the less massive companion (due to its larger surface area), the SPY `tail' distribution (consisting mainly of real DWDs) represents that of the secondary mass components. We thus experiment by replacing our default secondary mass distribution with that of the systems that are in the `tail' of the \drvm\ distribution of the SPY sample, using a three-component Gaussian mixture with $w = \{0.4, 0.35, 0.25\}, \mu = \{0.51, 0.35, 0.68\}$\,\msun\ and $\sigma = \{0.024, 0.087, 0.1\}$\,\msun\ \citep[][`Secondary SPY tail' in Fig.~\ref{fig:model_variations}]{maoz17}. In this case, we follow \citet{maoz17} and use a Gaussian with mean 0.75\,\msun\ and $\sigma=0.25$\,\msun\ for the primary mass distribution (similar to the photometric secondary mass distribution of \citealt{bro16}).

Third, we change the mass limit that defines ELM WDs from $0.25$\,\msun\, which is our default assumption, to $0.3$\,\msun (`ELM secondary flat' model variation), following \citet{bro10,bro16}. In addition to the ELM limit mass, we run a model variation in which we draw ELM companions from a normal distribution with a mean at 0.76\,\msun\ and a standard deviation of 0.25\,\msun\ that agrees better with the ELM observations \citep[][`ELM secondary Gaussian' model variation]{bro16}. 

Finally, we consider the most extreme combinations of $\alpha$ and $f_{\rm DWD}$ allowed within the $1\sigma$ contour reported by \citet{maoz18}. 
Specifically, we run two additional model variations with 1) $\alpha = -1.45, f_{\rm DWD} = 0.078$ and 2) $\alpha = -1.18, f_{\rm DWD} = 0.112$.

\section{Estimating the spectral shape of the unresolved foreground signal} \label{sec:gwg}

To make predictions of the spectral shape of the unresolved foreground signal for each of the proposed models, we use the iterative scheme presented in~\citet{kar21}. This scheme begins with the generation of the signal measured by \textit{LISA}, by computing the wave-forms for each of the simulated catalogue entries and by projecting it on the \textit{LISA} arms for a given duration of the mission. In our case we set the duration to $T_{\rm obs}=4\,$yr. At the time of writing, this figure is still under evaluation~\citep[e.g.][]{LISA_mission_duration}. Thus, our analysis do not account for the duty cycle of the detector and ignores possible realistic data analysis implications due to interruptions in the \textit{LISA} data stream~\citep[e.g.][]{bag19,dey21}. Nevertheless, for the case of continuous signals, such as those produced by DWDs, the number of resolved sources scales proportionally to their total in-band time after the first year of observation~\citep[e.g.][]{kar21,Cornish17}. Considering the above, the effect of a given duty cycle percentage has indeed an impact on our results here, but the figures presented in this work can be roughly scaled accordingly to the given total in-band time.

We employ the quasi-monochromatic wave-form described by eight parameters $\{f, \dot{f}, \cal{A}, \iota, \lambda, \beta, \psi, \phi_{\rm 0} \}$ (see Section~\ref{sec:pop}), neglecting external and/or environmental factors that may introduce a modulation of this signal \citep[e.g.][]{bar15,rob18,dan19,kor19}. This first step is naturally the most computationally expensive, because each catalogue contains $\sim 10^7$ entries. To ease the computational burden of the subsequent step of the analysis, we also compute for each of the wave-forms the optimal SNR in isolation $\rho^\mathrm{iso}_i$. The $\rho^\mathrm{iso}_i$ is the SNR of a given source $i$, in the absence of other signals, with respect to the instrumental noise only. The SNR of a given source is calculated as
\begin{equation}
\rho_\mathrm{tot}^2 = \sum_K  \left( h_K | h_K \right),
\label{eq:snrtot}
\end{equation}
with $K \in \{A, E \}$ the noise-orthogonal Time Delayed Interferometry (TDI) variables~\citep{tdi,aet}, 
and $\left( \cdot | \cdot \right)$ denotes the noise weighted inner product expressed for two time 
series $a$ and $b$, as
\begin{equation}
\left( a | b \right) = 2 \int\limits_0^\infty \mathrm{d}f \left[ \tilde{a}^\ast(f) \tilde{b}(f) + \tilde{a}(f) \tilde{b}^\ast(f) \right]/\tilde{S}_n(f).
\label{eq:ineerprod} 
\end{equation}

The second step of the analysis begins with an estimate of the unresolved signal Power Spectral Density (PSD), $S_{ \rm n,\,k}$, which is the sum of the gravitational-wave sources and the instrumental noise. Initially, we estimate the PSD by using a running median on the power spectrum of the data. Then, this initial estimate is smoothed further by either fitting polynomial models, performing spline fits, or by using appropriate smoothing kernels. Next, we adopt an SNR threshold ($\rho_0$), which we compare to the SNR ($\rho_i$) for each source $i$, using the smoothed $S_{\rm n,\,k}$ as the total noise PSD. If $\rho_i > \rho_0$, the source is considered detectable, and is subtracted from the data. To accelerate this procedure, we make use of the previously calculated $\rho^\mathrm{iso}_i$: if $\rho^\mathrm{iso}_i$ is too low, we can skip the wave-form calculation altogether. After  subtracting the brightest sources in this step, we re-evaluate the smoothed PSD of the residual $S_{\rm n,\,k+1}$ and repeat the procedure until convergence. Convergence is reached when no sources are found below the given $\rho_0$ threshold, or if the difference between $S_{\rm n,\,k+1}$ and $S_{\rm n,\,k}$ is negligible at all frequencies. 
At the end of this procedure, we compute the final SNR with respect to the final estimate of $S_{ \rm n,\,k_\mathrm{final}}$, for the sources recovered. At the same time, we perform a Fisher analysis in order to estimate the accuracy of the recovery of the parameters, again with respect to the final $S_{\rm n,\,k_\mathrm{final}}$. The Fisher Information Matrix (FIM) for each recovered source is calculated as 
\begin{equation}
	\left. F_{ij} = \left( \frac{\partial h(\vec{\theta})}{\partial \theta_i} \bigg| \frac{\partial h(\vec{\theta})}{\partial \theta_j}  \right) \right\vert_{\vec{\theta} = \vec{\theta}_\mathrm{true}},
	\label{eq:fim}
\end{equation}
where $h$ is the template of each signal, and $\vec{\theta}$ represents the wave-form parameter vector. The inverse of $F_{ij}$ yields the covariance matrix with the diagonal elements being the mean
square errors on each parameter $\theta_i$, and the off-diagonal elements describing the correlations between the parameters $\rho_{\theta_i \theta_j}$.
We remark that errors derived using FIM analysis are valid for relatively high SNRs \citep[e.g.][]{cut98,val08}. Thus, we expect that in some cases the errors may be underestimated. A full Bayesian parameter estimation is required to derive more realistic uncertainties \citep[e.g.][]{bus19,lit20,roe20}.

The pipeline described above has an important caveat: for each source that is being recovered, we assume that the parameters are perfectly estimated and that its waveform is perfectly subtracted from the data (i.e. leaving no residuals behind). This means that our estimate of the unresolved signal measured by \textit{LISA} may be optimistic. There have been proposals for data analysis pipelines~\cite{lit20}, aiming to disentangle the signals and recover a fraction of the sources, depending on the \textit{LISA} sensitivity. These methods are based on stochastic algorithms, which are in general computationally expensive.
The method employed in this study, on the other hand, allows the processing of catalogues consisting of millions of sources fast and at low computational costs, while at the same time it allows us to perform estimates on the parameter errors recovery employing a FIM approach.

\section{Results} \label{sec:results}

\begin{figure}
	\includegraphics[width=1\columnwidth]{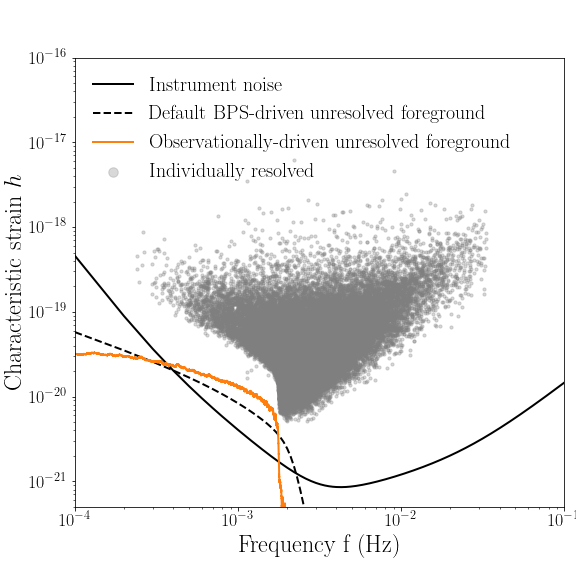}
	\caption{Characteristic strain--frequency plot: the solid black line shows the instrumental noise, grey points represent DWDs with SNR~$>5$ after 4\,yr of data acquisition with \textit{LISA}, while the solid orange line shows the unresolved confusion foreground produced by our observationally driven population. For comparison, the confusion foreground reported in the \textit{LISA} mission proposal produced using a BPS-based model \citep{lisa} is plotted as a dashed black line.}
    \label{fig:strain_plot}
\end{figure} 

\begin{figure*}
	\includegraphics[width=2\columnwidth]{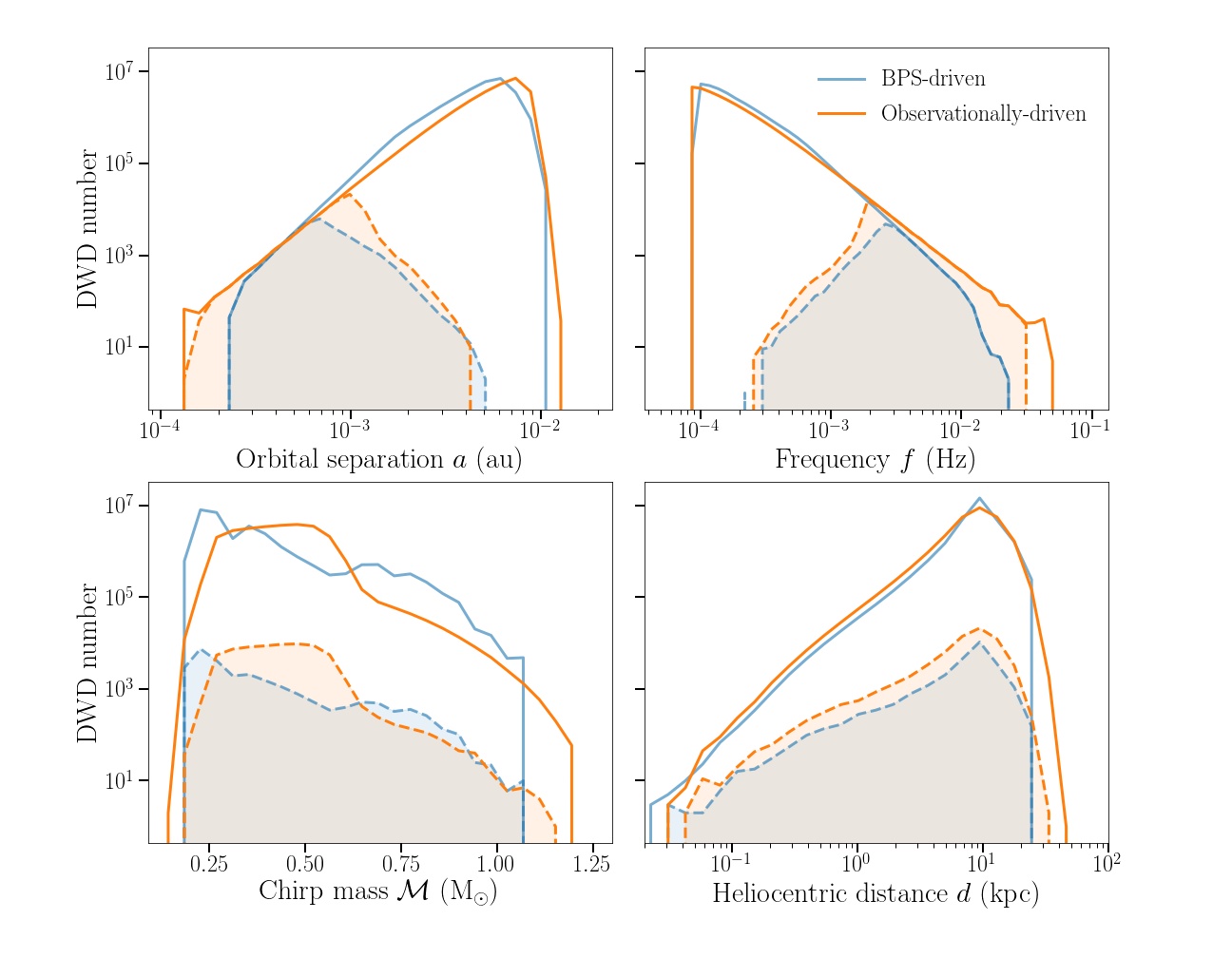}
    \caption{Distributions of the DWD orbital separations (top left), gravitational-wave frequencies (top right), chirp masses (bottom left) and distances (bottom right); in logarithmic intervals: the underlying mock population is represented by a solid line, while the sub-population of individually detected DWDs is delimited with a dashed line and is highlighted in colour. The observationally driven population is plotted in orange, while the BPS-driven population from \citet{kor17} is plotted in blue for comparison. Top panels of the figure highlight a strong selection effect with gravitational-wave frequency (orbital separation) for \textit{LISA}, as expected.}
    \label{fig:distributions}
\end{figure*} 

\begin{figure}
	\includegraphics[width=1\columnwidth]{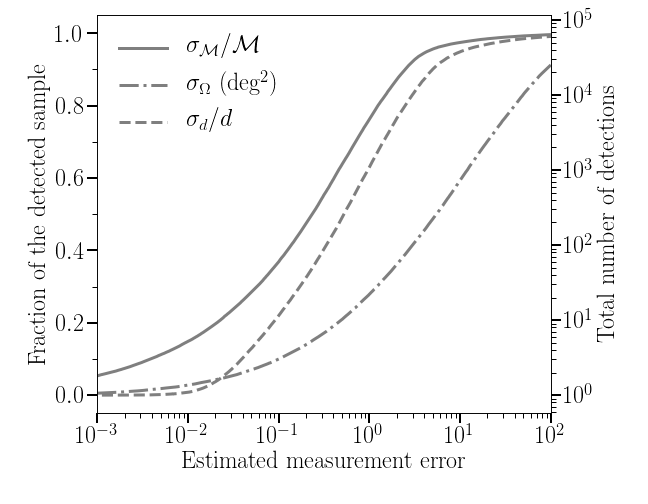}
    \caption{Cumulative distributions of the estimated \textit{LISA} fractional error on the chirp mass $\sigma_{\cal M}/{\cal M}$ (solid line), sky position error $\sigma_{\Omega}$ shown in deg$^2$ (dashed-dotted line) and distance $\sigma_d/d$ (dashed line). The secondary y-axis shows the corresponding total number of individual detections. We find that the fractional error on the frequency, $\sigma_f/f \ll 10^{-5}$ for all detected binaries, and, thus, it is not shown in this figure.}
    \label{fig:LISA_errors}
\end{figure} 

We feed our default DWD synthetic catalogue ($\alpha = -1.3$ and $f_{\rm DWD} = 0.0095$) into the pipeline described in Section~\ref{sec:gwg}, setting the mission lifetime to $T_{\rm obs}=4$\,yr, and adopting conservative SNR threshold to $\rho_0=7$. 

Our default model yields $\sim 60 \times 10^3$ detectable DWDs with SNR $>7$, which is higher than previous estimates.  In Fig.~\ref{fig:strain_plot} we show the sample of detected DWDs (grey points) and the residual spectrum after subtracting the detectable sources (solid orange line). The residual spectrum constitutes the unresolved confusion foreground from our simulated DWD population (see Section~\ref{sec:gwg}). For comparison, we also show the confusion foreground resulting from the BPS-based DWD population of \citet{kor17} employed in the \textit{LISA} mission proposal \citep{lisa}. We note that the shape of the two foreground signals is significantly different. The observationally driven foreground is higher at $f<1.5\,$mHz, then it sharply drops at 2\,mHz. In comparison, the BPS-based one extends up to almost 3\,mHz. Note that the exact extension of the DWD foreground is of particular relevance for the detectability of extreme mass ratio inspirals that are expected to be at the detection threshold at a few mHz frequencies \citep[e.g.][]{Babak2017, bon20}.

Distributions of the orbital separation, gravitational-wave frequency, chirp mass, and distances of detected DWDs (orange dashed) are compared to those of the underlying mock population (orange solid) in Fig.~\ref{fig:distributions}. By examining these two populations, it is evident that detected binaries trace the shape of the underlying `true' chirp mass and distance distributions. The top panels of Fig.~\ref{fig:distributions} reveal that \textit{LISA} can essentially detect all simulated DWDs with $f>1\,$mHz ($a < 1 \times 10^{-3}$\,au). We note that a small sub-sample of $\sim80$ DWDs with $f>33\,$mHz remains undetected in our simulation. 
This is an artefact of our data generation process. Specifically, it results from setting the time step to 15\,s during the data generation process. This choice implies that, when transforming DWD signals into the frequency domain, the maximum frequency of the data is $1/dt/2 = 33$\,mHz, according to the Nyquist–Shannon sampling theorem. Thus, any binary with a characteristic frequency higher than 33\,mHz is `lost' in our analysis. 
Finally, we note that at $f>1\,$mHz ($a < 1 \times 10^{-3}$\,au) the detected sample suffers from a selection bias due to the rapidly rising instrumental noise and the additional noise contribution from the unresolved DWDs (see Fig.~\ref{fig:strain_plot}).

Equations~\eqref{eq:fdot} and \eqref{eq:amp} imply that the chirp mass and the distance can be estimated if $f, {\cal A}$ and $\dot{f}$ are measured for a binary. 
In Fig.~\ref{fig:LISA_errors} shows the cumulative distribution of the estimated \textit{LISA} fractional errors on the chirp mass 
\begin{equation}
    \frac{\sigma_{\cal M}}{{\cal M}} = \sqrt{ \left( \frac{11}{5} \frac{\sigma_f}{f} \right)^2 + \left( \frac{3}{5} \frac{\sigma_{\dot{f}}}{\dot{f}} \right)^2 + \frac{33}{25} \left(\frac{\sigma_f}{f} \right) \left( \frac{\sigma_{\dot{f}}}{\dot{f}} \right) \rho_{f\dot{f}}},
\end{equation}
the luminosity distance 
\begin{multline}
    \frac{\sigma_d}{d} = \sqrt{ \left( \frac{\sigma_{\cal A}}{{\cal A}} \right)^2 + 9 \left( \frac{\sigma_f}{f} \right)^2 + \left( \frac{\sigma_{\dot{f}}}{\dot{f}} \right)^2 +6 \left( \frac{\sigma_f}{f} \right)  \left( \frac{\sigma_{{\cal A}}}{{\cal A}} \right) \rho_{f{\cal A}} }\\
    \textstyle
\overline{\rule{0pt}{17pt}
    + 2 \left( \frac{\sigma_{\cal A}}{{\cal A}} \right)  \left( \frac{\sigma_{\dot{f}}}{\dot{f}} \right) \rho_{{\cal A}\dot{f}}+ 
6 \left( \frac{\sigma_f}{f} \right)  \left( \frac{\sigma_{\dot{f}}}{\dot{f}} \right) \rho_{f\dot{f}}
  \ },
\end{multline}
and the absolute error on the sky position
\begin{equation}
    \sigma_\Omega = 2\pi \sigma_\lambda \sigma_\beta \sqrt{1-\rho^2_{\lambda \beta}},
\end{equation}
where $\sigma_{\cal A}$, $\sigma_{f}$, $\sigma_{\dot{f}}$, $\sigma_\lambda$ and $\sigma_\beta$ are the $1\sigma$ errors, while $\rho_{{\cal A}f}$, $\rho_{{\cal A} \dot{f}}$, $\rho_{f \dot{f}}$ and $\rho_{\lambda \beta}$ are the respective correlation coefficients \citep[see also figure~6 of][]{kar21}.
The $1\sigma$ errors and the correlations are estimated with the Fisher analysis briefly outlined in Section~\ref{sec:gwg}. The fractional errors on the frequency, which are $\ll 10^{-5}$ across the entire sample, are omitted from Fig.~\ref{fig:LISA_errors}. More quantitatively, we estimate that the chirp mass and the distance could be constrained ($\sigma_\theta/\theta<$30\,per cent) for 55\,per cent and 40\,per cent of the \textit{LISA's} sample, respectively. We also find that the sky position could be determined with accuracy better than 1\,deg$^2$, for about 30\,per cent of DWDs. Combined with the measurements of $f$ and $\dot{f}$, this will enable searches of electromagnetic counterparts based on the photometric variability \citep[eclipses, ellipsoidal modulation, and effects of companion irradiation, e.g.][]{Burdge_2019}.
Our results reinforce the potential of \textit{LISA} in providing large statistical samples of several thousand detected binaries for studying the overall properties of the Galactic DWD population and for mapping the Milky Way shape \citep[e.g.][]{ada12, kor19, kor21, wil21}.

\begin{figure*}
	\includegraphics[width=1.4\columnwidth]{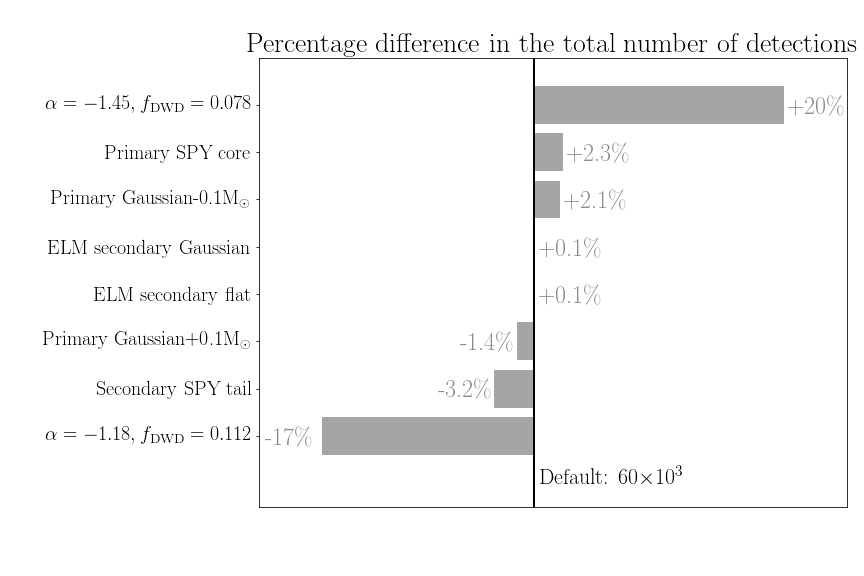}
    \caption{Difference in the number of \textit{LISA} detections for different model variations (see Section~\ref{sec:model_variations}) compared to the default model  ($\alpha = -1.3$ and $f_{\rm DWD} = 0.0095$), which produces $60 \times 10^3$ DWD with SNR$>7$ for a 4-year long \textit{LISA} mission.}
   \label{fig:model_variations}
\end{figure*}

Finally, we consider eight alternative model variations that affect the primary and secondary mass distributions, the mass limit for ELM WDs and the mass distribution of ELM companions, the power-law index of the DWD orbital separation distribution, $\alpha$, and the DWD fraction, $f_{\rm DWD}$ (see Section~\ref{sec:model_variations}). Fig.~\ref{fig:model_variations} summarises the obtained results in terms of the relative difference in the number of detectable binaries with SNR $>7$ compared to our default model ($60 \times 10^3$). We observe that the largest differences arise when changing the values of $\alpha$ and $f_{\rm DWD}$. In particular, lower values of $\alpha$ favour binaries with shorter periods, and, hence, produce more \textit{LISA} detections. Indeed, $f_{\rm DWD}$ within the \textit{LISA}'s frequency band increases from 0.95\,per cent (default) to 1.06\, per cent (for $\alpha =-1.45$), and decreases to 0.85\,per cent (for $\alpha =-1.18$).
Models assuming different distributions for the primary WD mass produce only moderate variations in the number of (individually) detected systems, of a few per cent. Finally, the alternative ELM models considered here have a negligible effect on the results.

\section{Discussion and conclusions} \label{sec:discussion}

In this study, we generated an observationally driven population of Galactic DWDs for the \textit{LISA} mission. We assembled this population following the model of \citet{maoz18}, constructed to fit the \drvm\ distribution obtained by combining the SDSS and SPY multi-epoch spectroscopic samples. Specifically, we assumed a constant star formation rate, a DWD fraction of $f_{\rm DWD} \simeq 1\,$per cent (valid within the \textit{LISA} band), a power-law orbital separation distribution at the DWD formation with an index $\alpha = -1.3$, and that the primary masses follow the mass distribution of single WDs from \citet{kep15}. Note that assumptions on the $f_{\rm DWD}$, the orbital separation, and the chirp mass distributions represent the main differences compared to the previous \textit{LISA}-focused studies based on BPS simulations. 

Our study identified two main results: the observationally driven model yields 1) a factor $2 - 5$ increase in the number of detectable DWDs compared to the BPS-based forecasts \citep[e.g.][]{kor17,lam19,bre20,wil21} and 2) a significantly different shape of the DWD confusion foreground (see Fig.~\ref{fig:strain_plot}). A comparison between several model alternatives, which we constructed by modifying some of the underlying assumptions, shows that our observationally driven model is robust, and that the uncertainty of the total number of detected sources is of 20\,per cent. When comparing, for example, to BPS-driven population constructed for \textit{LISA} by \citet{kor17}, we find that the observationally driven model generates 2.4 times more \textit{LISA} detections despite a similar total number of DWDs that both models predict in the \textit{LISA} frequency band ($26 \times 10^6$ DWDs for both models). When comparing with other recent BPS-driven forecasts \citep[e.g. the most recent ones by][]{lam19,bre20} the difference in the number of detected sources is even larger, although the same \textit{LISA} noise curve and mission duration of 4\,yr are employed in those studies. 

Figure~\ref{fig:distributions} illustrates the differences between our observationally driven population (orange) and a BPS-driven model (blue) of \citet{kor17}. It reveals that the reason for the enhanced number of \textit{LISA} detections in our observationally driven model is twofold. Firstly, our model presents an excess of binaries at high frequencies ($f>$~a few mHz), all of which are detectable by \textit{LISA}.
Secondly, DWDs in the observationally driven population on average have larger chirp masses and, thus, larger GW amplitudes, which yields more detectable DWDs and produces a sharp drop in the unresolved foreground at $f \sim 2$\,mHz. However, the BPS-based model provides more binaries with ${\cal M}>0.6\,$M$_\odot$ at $f<2\,$mHz, which helps to lower the unresolved foreground level at $f<2\,$mHz (see Fig.~\ref{fig:strain_plot}). Note that the two models assume the same spatial distribution (Eq.~\ref{eqn:dwd_positions}), and have been processed using the same \textit{LISA} data analysis pipeline (Section~\ref{sec:gwg}), therefore the DWD frequencies and the chirp masses are the main quantities driving the differences.

We highlight that it is particularly hard to constrain the DWD chirp-mass distribution with currently available observations. This is due to the many biases in the known DWD sample, and to the fact that only lower limit estimates are available for the masses of most of the photometric secondaries, which often remain unseen (see Fig.~\ref{fig:KnownDWDs} and Section~\ref{sec:obs}). Consequently, testing BPS-based predictions against observations in terms of WD masses or mass combinations such as mass ratio or chirp mass is also difficult. Based on the observed DWD systems known at that time, \citet{nel00} determined the possible masses and radii of the DWD progenitors by fitting detailed stellar evolution models, and used these to reconstruct the past mass-transfer phases. This led to an important insight that the standard common-envelope formalism (the $\alpha$-formalism, see \citealt{iva13} for details)---equating the energy balance in the system and implicitly assuming angular momentum conservation---does not work for the first phase of mass transfer in the DWD evolution. They concluded that studies that only adopt the $\alpha$-formalism will not reproduce the observed mass ratios of DWDs, which are close to 1, and that an alternative formalism for the first phase of the mass-transfer is required. This result was later confirmed with larger samples \citep{nel05,slu06}, and with the BPS approach as well \citep{nel01a, too12}. There is also a strong indication for a nearly-equal-mass `twin' excess in main-sequence binaries \citep{Moe_2017, ElBadry_2019}. However, it is still unclear how this excess affects the later DWD stage, following up to four mass transfer or common-envelope episodes. Interestingly, the SPY sample has a relatively large fraction of double-lined systems among the DWD candidates \citep[][Hallakoun et al., in preparation]{nap20}. This could hint at a possible twin excess also in DWD systems, since double-lined DWDs have mass ratios very close to 1 (otherwise the less massive WD outshines its more massive companion). However, this result should be corrected for the observational bias that favours the detection of double-lined systems, that are more luminous than single-lined systems, and can be identified using a single spectrum, regardless of the RV measurement uncertainty (as long as the spectral line cores are resolved). Therefore, given that the observational constraints are still poor, the WD mass distributions assumed in this study are as plausible as those emerging from BPS simulations. We demonstrated that changing our assumption on the WD masses using suggested distributions from different observational studies has only a moderate effect on the number of detected DWDs (see Fig.~\ref{fig:model_variations}).

Our assumption on the orbital separation/frequency distribution is based on the tight constraint derived in \citet{maoz18}. We note that the analysis of \citet{maoz18} assumes a continuous power-law for the separation distribution up to separations of $\sim 4$\,au. BPS results indicate a possible gap in the orbital separation distribution between $\sim 0.2-1$\,au \citep{too14}. If the actual separation distribution truncates around $\sim 0.2$\,au, the DWD fraction derived from the results of \citet{maoz18} for the \textit{LISA} band could be up to two times higher than estimated here.Our assumption on the DWD orbital separation implies that different types of DWDs are born from the same power-law distribution. We argue that in the range accessible to \textit{LISA}, the shape of the separation distribution is heavily dominated by gravitational-wave radiation rather than by binary evolution for both BPS- and observation-based models (see Fig.~\ref{fig:distributions}, and see also fig.~7 of \citealt{lam19}), and thus does not affect our results. However, when modelling DWDs at larger orbital separation this simplification should be address in more detail.

It is also worth noting that the total number of DWDs in our observationally driven model is set by the observed local WD space density, $\rho_{\rm WD,\odot}$, and DWD fraction, $f_{\rm DWD}$. Current uncertainties on these two quantities affect the total number of DWD in the \textit{LISA} band (see Section~\ref{sec:DWD_tot}), and, consequently, the number of \textit{LISA} detections, by about 20\,per cent. We also note that there is a strong anti-correlation between the close binary fraction and the metallicity in main-sequence stars \citep{Moe_2019}, that might affect the DWD formation efficiency as a function of age. \citet{thi21} show that when implementing this anti-correlation in BPS-based DWD models, the size of the \textit{LISA}-detectable sample decreases by a factor of 2. The quantities discussed above may contribute to the discrepancy in the predicted number of \textit{LISA} detections between BPS-driven estimates and our study.

Although the observationally motivated constraints used here represent one of the main strengths of our study, they also introduce some limitations. One of the most obvious limitations is given by the accessible volume for WDs with SDSS and SPY, which is limited to a few kpc at most. In general, completeness ($>98\,$per cent) is currently achieved only up to $20-40$\,pc \citep[e.g.][and almost up to 100\,pc, as demonstrated by \citealt{jim18}]{hol2018,mcc20,gen21}. Thus, when constructing the population for \textit{LISA}, that can detect DWDs in the entire Milky Way, including its satellites \citep[e.g.][]{kor18}, we extrapolate local estimates of the WD space density, DWD fraction, and properties of DWDs and WDs in general, to the whole Galaxy. In addition, the observed samples are affected by non-trivial selection effects and biases that are hard to account for. For instance, both SPY and SDSS, being magnitude-limited samples, are likely to be biased towards low-mass WDs.

In this study, we used the DWD fraction derived in \citet{maoz18} for DWD orbital separations of $\lesssim 4\,$au, based on the combined SDSS and SPY samples. 
More recently, \citet{nap20} estimated a binary fraction of $\sim6$\,per cent based on a $\chi^2$ test assuming a single WD, using a larger sub-sample of $625$ WDs from SPY. This apparent inconsistency with the results of \citet{maoz17} is reconciled when taking into account the relevant orbital separation range: the detection efficiency of \citet[][fig.~6]{nap20} drops sharply for orbital periods longer than $\sim 10$\,d, corresponding to an orbital separation of $\sim 0.1$\,au, while \citet{maoz17} estimated the binary fraction within $\lesssim 4$\,au. If we scale down the binary fraction of \citet{maoz17} to the separation range $\lesssim 0.1$\,au, the $\sim6$\,per cent fraction is recovered (Eq.~\ref{eq:scaled_fbin}).
In addition, \citet{nap20} make two further claims against the \drvm\ method of \citet{maoz17}. First, they claim that the \drvm\ method does not account for the number of spectra available for each system. This is not the case since during the generation of the synthetic DWD population, each theoretical RV curve is sampled according to a sampling sequence drawn from the observed sample. Second, \citet{nap20} argue that the \drvm\ method ignores the varying statistical uncertainties between spectra. However, the synthetic RV measurements of the \drvm\ method are applied measurement error drawn from a Gaussian distribution similar to that of the observed sample.
Nevertheless, we note that a preliminary analysis of the ongoing follow-up effort of the DWD candidates from the SDSS and SPY samples (Hallakoun et al., in preparation) suggests that the actual DWD fraction within 4\,au might be somewhat lower than the $\sim 10$\,per cent estimated in \citet{maoz18}.

Studies of the local 20\,pc WD sample estimate a DWD fraction of $\lesssim 5$~per cent \citep{too17, hol2018}, consistent with the BPS predictions of \citet{too17}. This estimate is based on one confirmed DWD system, and another six DWD candidates selected due to their low mass, which cannot be formed with single-star evolution. It is possible, however, that more massive DWDs are still hidden in the 20\,pc sample. \citet{bel20} estimated DWD fraction of $2.5$\,per cent based on the astrometric wobble in the \textit{Gaia} data. Note, however, that their result relates to a different binary separation range. They may not be sensitive to some of the systems, depending on the mass ratio, luminosity ratio, and distance. In any case, the results presented in this study can be linearly re-scaled to any desirable $f_{\rm DWD}$ value as it enters only in the estimate of the total number of DWDs (see Section~\ref{sec:DWD_tot}).

Our DWD population has been constructed for the Milky Way disc ignoring the bulge/bar and the stellar halo, since some inherent assumptions from \citet{maoz18} (e.g. constant star formation rate) are not appropriate elsewhere. Thus, our results should be considered as a lower limit.
Based on the stellar mass, DWDs in the bulge/bar and the stellar halo are expected to be sub-dominant contributors to the \textit{LISA}-detectable population. For instance, given that the ratio of stellar mass in the bulge/bar region is $\sim0.3$ of that in the disc \citep{bla16}, we expect that accounting for the bulge/bar mass could boost the number of \textit{LISA} detections by up to 30\,per cent. Note, however, that since in this simple estimate we do not account for the star formation history of the bulge/bar, it should be considered as a generous upper limit. The mass of the stellar halo is only $\sim 0.01$ of the disc stellar mass, meaning that its contribution to the number of detections can be neglected \citep[see also a dedicated study by][]{rui09}. 

For the density distribution of DWDs in the disc, we employed an exponential radial profile with an isothermal vertical distribution (Eq.~\ref{eqn:dwd_positions}), which has been often employed in the BPS-driven \textit{LISA} studies. This allowed us to compare the number of detected DWDs with those studies directly \citep{kor17}. 
Any changes to Eq.~\ref{eqn:dwd_positions}---value of the scale parameters ($R_{\rm d}$ and $z_{\rm d}$) or its functional form---would affect the volume of the Galaxy, and, hence, the total number of DWDs (see Section~\ref{sec:DWD_tot}). 
Our results for detectable binaries can be linearly re-scaled to the volume corresponding to any desired set of $R_{\rm d}$, $R_\odot$ and $z_{\rm d}$. We note, though, that these parameters may also change the shape of the confusion foreground.
\citet{ben06} explored the effect of changing $z_{\rm d}$ on the Galactic confusion foreground. They demonstrated that at a fixed space density, an increase in $z_{\rm d}$ (leading to an increase in the total number of DWDs) raises the overall confusion level at $f\lesssim 1-3$\,mHz and shifts the transfer frequency---i.e. the frequency above which it is possible to resolve all DWDs---to slightly higher values. They also showed that at a fixed total number of binaries, changing $z_{\rm d}$ does not influence the shape of the confusion foreground. We defer similar investigations to future publications.

Until very recently, the known sample of WDs was compiled of WDs discovered using different, limited, methods by various surveys, mostly by chance. This incomplete sample suffered from problematic and largely unknown selection biases that limited any statistical inference \citep[see][and references therein]{GentileFusillo_2019}. This has changed with the second data release of the \textit{Gaia} mission \citep{Gaia_2016, Gaia_2018}, which has revealed, for the first time, the nearby Galactic WD population in the least biased form so far \citep{GentileFusillo_2019}. The fifth generation of the SDSS \citep[SDSS-V;][]{Kollmeier_2017}, which has already started collecting data in 2020 \citep[with a first DWD was already published by][]{ved21}, will acquire multi-epoch spectra for a large fraction of the \textit{Gaia} WDs, revealing the properties of the Galactic WD population clearer than ever before.
In the nearer future, the follow-up study of the DWD candidates from the SDSS and SPY, should provide an updated value for the DWD fraction, as well as observational orbital separation and component-mass distributions, for the first time for the general DWD population (Hallakoun et al., in preparation).
In addition, the already prolific ZTF \citep{kup21} survey, and upcoming large-scale optical surveys such as BlackGEM \citep{bloem15}, GOTO \citep{ste17}, and the Vera Rubin Observatory \citep{LSST}, are expected to deliver a more unbiased sample across both hemispheres and at low Galactic latitudes, ahead of the \textit{LISA} launch.With the increasing sample size we expect the observational constraints to become tighter and more reliable. Any changes coming with new observations can be easily accounted for by re-scaling, or by directly implementing them in the model.
Therefore, the observationally driven DWD model presented here may be a handy tool for generating DWD populations for the space-base gravitational-wave observatories as an alternative to BPS-driven studies.
Finally, our results demonstrate the impact of the underlying DWD population on achieving the science objectives of the \textit{LISA} mission, and the need for an observationally driven characterisation of the DWD population.

We thank Carles Badenes, Sihao Cheng, Dan Maoz, Tom Marsh, Riccardo Buscicchio, Eliot Finch, Davide Gerosa, Antoine Klein, Mauro Pieroni, Christopher J. Moore, Alberto Vecchio, and the anonymous referee for useful suggestions and discussions.
VK and ST acknowledge support from the Netherlands Research Council NWO (respectively Rubicon 019.183EN.015 and VENI 639.041.645 grants).
The research of NH is supported by a Benoziyo prize postdoctoral fellowship. NK acknowledges the support from the ESA Prodex funding program (Gr-Prodex-Call 2019).

This research made use of the tools provided by the \textit{LISA} Data Processing Group (LDPG) and the \textit{LISA} Consortium \textit{LISA} Data Challenges (LDC) working group\footnote{\href{https://lisa-ldc.lal.in2p3.fr/}{https://lisa-ldc.lal.in2p3.fr/}}.

This research made use of \textsc{astropy}\footnote{\href{http://www.astropy.org}{http://www.astropy.org}}, a community-developed core \textsc{python} package for Astronomy \citep{Astropy_2013, Astropy_2018}, \textsc{matplotlib} \citep{Hunter_2007}, \textsc{numpy} \citep{Numpy_2006, Numpy_2011}, and \textsc{scipy} \citep{Virtanen_2020}.

%%%%%%%%%%%%%%%%%%%%%%%%%%%%%%%%%%%%%%%%%%%%%%%%%%
\section*{Data Availability}
This study is based on published results. 
The \textsc{python} code for generating an observationally driven DWD population for \textit{LISA} is available at \url{https://gitlab.in2p3.fr/korol/observationally-driven-population-of-galactic-binaries}. Generated data have been processed using the pipeline presented in \citet{kar21} and available at \url{https://gitlab.in2p3.fr/Nikos/gwg}.



\bibliographystyle{mnras}
\bibliography{DWDpop} % if your bibtex file is called example.bib

%\begin{thebibliography}{99}
%\end{thebibliography}

%%%%%%%%%%%%%%%%%%%%%%%%%%%%%%%%%%%%%%%%%%%%%%%%%%

%%%%%%%%%%%%%%%%% APPENDICES %%%%%%%%%%%%%%%%%%%%%

%\appendix


%If you want to present additional material which would interrupt the flow of the main paper, it can be placed in an Appendix which appears after the list of references.

%%%%%%%%%%%%%%%%%%%%%%%%%%%%%%%%%%%%%%%%%%%%%%%%%%

% Don't change these lines
\bsp	% typesetting comment
\label{lastpage}
\end{document}